\documentclass[aps,twocolumn,showpacs,preprintnumbers,amsmath,amssymb,superscriptaddress,floatfix,nofootinbib]{revtex4}

\usepackage{mathrsfs,bm}
\usepackage{longtable,lscape}
\usepackage{txfonts}
\usepackage{amssymb}
\usepackage{indentfirst}
\usepackage{graphicx,,booktabs}
\usepackage{multirow}
\usepackage{color}
\usepackage{amssymb}
\usepackage{subfigure}

\newcommand{\Slash}[1]{\ooalign{\hfil/\hfil\crcr$#1$}}

\begin{document}

\title{The role of $a_1(1260)$ in $\pi^- p \to a^-_1(1260) p$ and $\pi^- p \to \pi^- \rho^0 p$ reactions near threshold}

\author{Chen Cheng}
\affiliation{Institute of Modern Physics, Chinese Academy of
Sciences, Lanzhou 730000, China} \affiliation{University of Chinese
Academy of Sciences, Beijing 100049, China}

\author{Ju-Jun Xie}
\email{xiejujun@impcas.ac.cn} \affiliation{Institute of Modern
Physics, Chinese Academy of Sciences, Lanzhou 730000, China}
\affiliation{University of Chinese Academy of Sciences, Beijing
100049, China} \affiliation{Research Center for Hadron and CSR
Physics, Institute of Modern Physics of CAS and Lanzhou University,
Lanzhou 730000, China}

\author{Xu Cao}
\affiliation{Institute of Modern Physics, Chinese Academy of
Sciences, Lanzhou 730000, China} \affiliation{University of Chinese
Academy of Sciences, Beijing 100049, China} \affiliation{Research
Center for Hadron and CSR Physics, Institute of Modern Physics of
CAS and Lanzhou University, Lanzhou 730000, China}

\date{\today}

\begin{abstract}

We report on a theoretical study of the $\pi^- p \to a^-_1(1260) p$
and $\pi^- p \to \pi^- \rho^0 p$ reactions near threshold within an
effective Lagrangian approach. The production process is described
by $t$-channel $\rho^0$ meson exchange. For the $\pi^- p \to \pi^-
\rho^0 p$ reaction, the final $\pi^- \rho^0$ results from the decay
of the $a_1(1260)$ resonance which is assumed as a dynamically
generated state from the $K^* \bar K$ and $\rho \pi$ coupled channel
interactions. We calculate the total cross section of the $\pi^- p
\to a^-_1(1260) p$ reaction. It is shown that, with the coupling
constant of the $a_1(1260)$ to $\rho \pi$ channel obtained from the
chiral unitary theory and a cut off parameter $\Lambda_\rho \sim
1.5$ GeV in the form factors, the experimental measurement can be
reproduced. Furthermore, the total and differential cross sections
of $\pi^- p \to a^-_1(1260) p \to \pi^- \rho^0 p$ reaction are
evaluated, and it is expected that our model calculations can be
tested by future experiments. These reactions are important for the
study of the $a_1(1260)$ resonance and would provide further clue
for the nature of $a_1(1260)$ state.

\end{abstract}

\maketitle

\section{INTRODUCTION}

Within the picture of the classical quark model, the mesons are
bound states of quarks and antiquarks. This picture is very
successful. Most of the known mesons can be described very well
within the quark model~\cite{Agashe:2014kda}. However, it seems that
the meson spectrum is much richer than that predicted by the quark
model. There is a growing set of experimental observations of
resonance-like structures with quantum numbers which are forbidden
for the quark-antiquark system or situated at masses which cannot be
explained by the quark-antiquark
model~\cite{Klempt:2007cp,Brambilla:2014jmp}. For example, the new
observations~\cite{Choi:2003ue,Acosta:2003zx,Abazov:2004kp,Ablikim:2013mio,Liu:2013dau,Adolph:2015pws,D0:2016mwd}
have challenged the conventional wisdom that mesons are made of
quark-antiquark pairs in the low energy region.

In the quark model, the ground state axial-vector resonances are
$a_1(1260)$ and $b_1(1235)$ with $I^G(J^{PC}) = 1^-(1^{++})$ and
$1^+(1^{+-})$, respectively. The experimental mass $M_{a_1(1260)} =
1230 \pm 40$ MeV is more precisely than its width
$\Gamma_{a_1(1260)} = 250 \sim 600$ MeV assigned by the Particle
Data Group~\cite{Agashe:2014kda}. A recent COMPASS measurement
published in 2010~\cite{Alekseev:2009aa} provides a much smaller
uncertainty of the width of $a_1(1260)$ $\Gamma_{a_1(1260)} = 367
\pm 9 ^{+28} _{-25}$ MeV.

In the chiral unitary approach, the low lying axial-vector mesons,
$a_1(1260)$ and $b_1(1235)$, are composite particles of a vector
meson and a pseudoscalar meson in coupled
channels~\cite{Roca:2005nm}. Indeed, the $a_1(1260)$ is dynamically
generated from the $K^*\bar{K}$ and $\rho \pi$ channels and the
couplings of the $a_1(1260)$ to these channels can be also obtained
at the same time~\cite{Roca:2005nm}. Based on these results, the
radiative decay of the $a_1(1260)$ axial-vector meson was studied in
Refs.~\cite{Roca:2006am,Nagahiro:2008cv}, where the theoretical
calculations agree with the experimental values within
uncertainties.

Recently the COMPASS collaboration~\cite{Adolph:2015pws} reported
the observation of a resonancelike structure around $1.42$ GeV with
axial-vector quantum numbers $1^-(1^{++})$ in the $f_0(980) \pi$
$P$-wave of the $\pi^- \pi^- \pi^+$ final state, and it was claimed
as a signal as a new resonances that was named the $``a_1(1420)"$
state with width around $140$ MeV. It is very difficult to explain
this structure as a new state within the quark model, because the
radial excitation of $a_1(1260)$ is expected to have a mass above
$1650$ MeV. Furthermore, it is not expected that the radial
excitation state has a width which is much smaller than the one of
the ground state. In Refs.~\cite{Ketzer:2015tqa,Aceti:2016yeb}, the
$``a_1(1420)"$ state can be explained as a triangle singularity via
the decay of $a_1(1260)$ into $K^* \bar{K}$ and subsequent
rescattering of the $K$ from the $K^*$ decay to form the $f_0(980)$
resonance. In Ref.~\cite{Wang:2015cis}, the production of $a_1$
states are studied in heavy meson decays which can also provide
insights to the $a_1(1420)$ and the future experimental analyses
will very probably lead to a deeper understanding of the nature of
the $a_1(1420)$.

In this work, with the coupling of $a_1(1260)$ to the $\rho \pi$
channel which was obtained within the picture that the $a_1(1260)$
resonance is dynamically generated from the $K^* \bar{K}$ and $\rho
\pi$ channels~\cite{Roca:2005nm}, we study the role of $a_1(1260)$
resonance in the $\pi^- p \to a_1^-(1260)p$ and $\pi^- p \to \rho^0
\pi^- p$ reactions near threshold using an effective Lagrangian
approach. In our calculation, the $t$ channel $\rho^0$ exchange is
considered. The total cross sections of $\pi^- p \to a^-_1(1260) p$
reactio are calculated. It is found that the theoretical
calculations for the total cross sections of $\pi^- p \to
a^-_1(1260) p$ reaction are in agreement with the experimental data.
In addition, the total and differential cross sections for the
$\pi^- p \to a^-_1(1260) p \to \rho^0 \pi^- p$ reaction are
predicted and could be tested by future experiments. Because the
main decay channel of $a_1(1260)$ resonance is the $\rho \pi$
channel, the $\pi^- p \to a^-_1(1260) p \to \rho^0 \pi^- p$ reaction
is very useful to deep understanding the nature of $a_1(1260)$ state
and also the nature of the $a_1(1420)$.

This paper is organized as follows. In Sec.~\ref{sec:formalism},
formalism and ingredients used in the calculation are given. In
Sec.~\ref{sec:results}, the results are presented and discussed.
Finally, a short summary is given in the last section.

\section{FORMALISM AND INGREDIENTS} \label{sec:formalism}

The combination of effective Lagrangian method and isobar model is
an important theoretical approach in describing the meson production
processes. In this section, we introduce the theoretical formalism
and ingredients to calculate the $a_1(1260)$ hadronic production in
$\pi^- p \to a_1^-(1260)p$ and $\pi^- p \to \rho^0 \pi^- p$
reactions within the effective Lagrangian method.

\subsection{Feynman diagrams and interaction Lagrangian densities} \label{feylag}

\begin{figure}[htbp]
\begin{center}
\includegraphics[scale=0.65]{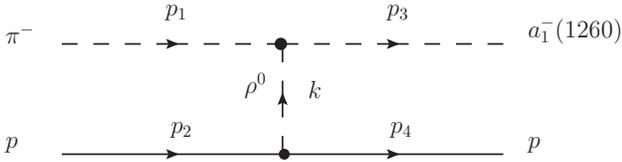}
\caption{Feynman diagrams for $\pi^- p \to a_1^-(1260)p$ reaction.
We also show the definition of the kinematical ($p_1$, $p_2$, $p_3$,
$p_4$, and $k$) that we use in the present calculation. In addition,
we use $k = p_2 - p_4$.} \label{piptoalpdiagram}
\end{center}
\end{figure}

\begin{figure}[htbp]
\begin{center}
\includegraphics[scale=0.6]{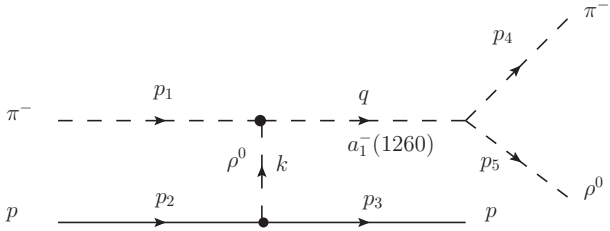}
\caption{Feynman diagrams for $\pi^- p \to \rho^0 \pi^- p$
reaction.} \label{piptopirhopdiagram}
\end{center}
\end{figure}

The basic tree level Feynman diagrams for the $\pi^- p \to
a_1^-(1260)p$ and $\pi^- p \to \rho^0 \pi^- p$ reactions are
depicted in Fig.~\ref{piptoalpdiagram} and
Fig.~\ref{piptopirhopdiagram}, respectively. For these reactions,
the $t$-channel $\rho^0$ exchange is considered in this calculation,
since the main decay channel of $a_1(1260)$ is the $\rho \pi$
channel.

To compute the contributions of diagrams shown in
Figs.~\ref{piptoalpdiagram} and \ref{piptopirhopdiagram}, we use the
interaction Lagrangian density for the $\rho NN$ vertex as in
Refs.~\cite{Machleidt:1987hj,Machleidt:1989tm,Doring:2010ap,Ronchen:2012eg},
\begin{eqnarray}
{\cal L}_{\rho NN} = -g_{\rho NN}\bar{N}(\gamma^\mu
-\frac{\kappa_{\rho}}{2m_N}\sigma^{\mu\nu}\partial_\nu)\vec{\tau}\cdot\vec{\rho}_{\mu}
N,
\end{eqnarray}
where the parameters are taken as commonly used
ones~\cite{Tsushima:2000hs,Gasparyan:2003fp,Xie:2007vs,Xie:2007qt,Xie:2015wja}:
$g_{\rho NN}=3.36$ and $\kappa_\rho=6.1$.

In addition, we need also the effective interaction of the
$a_1(1260)\rho \pi$ vertex. As mentioned before, in the chiral
unitary approach of Ref.~\cite{Roca:2005nm}, the $a_1(1260)$
resonance is dynamically generated from the interaction of $K^*
\bar{K}$ and $\rho \pi$ interactions. One can write down the $\pi^-
\rho^0 a^-_1(1260)$ vertex of Figs.~\ref{piptoalpdiagram} and
\ref{piptopirhopdiagram} as,
\begin{eqnarray}
\upsilon =
\frac{1}{\sqrt{2}}g_{a_1\rho\pi}\varepsilon^\mu(\rho)\varepsilon^*_\mu(a_1),\label{vertex}
\end{eqnarray}
where $\varepsilon^\mu(\rho)$ and $\varepsilon^*_\mu(a_1)$ are the
polarization vectors of $\rho$ and $a_1(1260)$. The $g_{a_1\rho\pi}$
is the coupling constant of the $a_1(1260)$ to the $\rho\pi$
channel, which is taken to be $(-3795, 2330)$ MeV as obtained in
Ref.~\cite{Roca:2005nm}. The factor $\frac{1}{\sqrt{2}}$ in
Eq.~\eqref{vertex} accounts for the fact that in the $I=1$ and $I_3
= -1$ combination of $\rho\pi$ mesons,
\begin{align}
|\rho\pi> {(I=1,\
I_3=-1)}=\frac{1}{\sqrt{2}}(\rho^0\pi^--\rho^-\pi^0).
\end{align}

With the effective Lagrangian densities above, we can
straightforwardly construct the invariant scattering amplitude for
$\pi^- p \to a^-_1(1260) p$ reaction corresponding to the Feynman
diagram in Fig.~\ref{piptoalpdiagram}:
\begin{eqnarray}
&& {\cal M} (\pi^- p \to a^-_1(1260) p) = -\frac{ig_{\rho NN}
g_{a_1\rho\pi} F_\rho(k) }{\sqrt{2}}
\bar u(p_4, s_4) \nonumber \\
&&\times [\gamma^\mu + \frac{\kappa_\rho}{2m_N}(k^{\mu} - \Slash k
\gamma^{\mu} )] u(p_2, s_2)
G_{\mu\nu}(k)\varepsilon^{\ast\nu}(p_3,s_3), \label{amplitude2}
\end{eqnarray}
where $s_4$, $s_2$ and $s_3$ are the polarization variables of final
proton, initial proton and $a_1(1260)$ resonance, respectively. The
$\rho$-meson propagator $G_{\mu\nu}(k)$ is,
\begin{align}
G_{\mu\nu}(k) &= -i\frac{g_{\mu\nu} - k_\mu k_\nu/m_{\rho}^2}{k^2 -
m_{\rho}^2},
\end{align}
where $m_{\rho}$ is the mass of the $\rho$ meson and we take $m_\rho
= 775.26$ MeV.

In Eq.~\eqref{amplitude2}, $F_{\rho}(k)$ is the form factor for
$\rho NN$ vertex and we take it as in
Refs.~\cite{Machleidt:1987hj,Machleidt:1989tm},

\begin{align}
F_\rho(k)
&=\left(\frac{\Lambda_{\rho}^2-m_{\rho}^2}{\Lambda_{\rho}^2 -
k^2}\right)^2, \label{rhoNNformfactor}
\end{align}
with $\Lambda_\rho$ the cut off parameter which will be discussed in
the following.

Similarly, we can get the invariant scattering amplitude for $\pi^-
p \to \pi^- \rho^0 p$ reaction corresponding to the Feynman diagram
in Fig.~\ref{piptopirhopdiagram}:
\begin{align}
& {\cal M}(\pi^-
p \to \pi^- \rho^0 p) = -\frac{ig_{\rho NN}g_{a_1\rho\pi}^2F_\rho(k)F_{a_1}(q)}{\sqrt{2}}\bar u(p_3, s_3) \notag \\
&\times [\gamma^\mu + \frac{\kappa_\rho}{2m_N}(k^{\mu} - \Slash k
\gamma^{\mu} )]u(p_2, s_2) G_{\mu\nu}(k) G^{\nu\sigma}(q)
\varepsilon^{*}_\sigma(p_5,s_5),  \label{amplitude3}
\end{align}
where $s_5$ is the polarization variable of $\rho^0$ meson, and
$G^{\nu\sigma}(q)$ is the $a_1(1260)$ propagator, which is,
\begin{align}
G^{\nu\sigma}(q)&=-i\frac{g^{\nu\sigma}-q^\nu
q^\sigma/m_{a_1}^2}{q^2-m_{a_1}^2+im_{a_1}\Gamma_{a_1}},
\end{align}
where $\Gamma_{a_1}$ and $m_{a_1}$ are the width and mass of the
$a_1(1260)$ resonance, respectively. We take $m_{a_1} = 1230$ MeV.
For $\Gamma_{a_1}$, as mentioned above, since its value has large
uncertainties, we take $\Gamma_{a_1} = 250$, $425$, and $600$ MeV
for comparison.

In Eq.~\eqref{amplitude3}, $F_{a_1}(q)$ is the form factor of
$a_1(1260)$ state. In our present calculation, we adopt the
following form as in many previous
works~\cite{Tsushima:2000hs,Gasparyan:2003fp,Xie:2007vs,Xie:2007qt,Xie:2015wja}:
\begin{align}
F_{a_1}(q)&=\frac{\Lambda_{a_1}^4}{\Lambda_{a_1}^4+(q^2-m_{a_1}^2)^2},
\end{align}
where $\Lambda_{a_1}$ is the cutoff parameter of $a_1(1260)$
resonance.

The differential cross section in the center of mass frame (c.m.)
for the $\pi^- p \to a_1^-(1260)p$ and $\pi^- p \to \rho^0 \pi^- p$
reactions can be derived from the invariant scattering amplitude
square $|{\cal M}|^2$, reading as:
\begin{align}
& \frac{d\sigma(\pi^- p \to a_1^-(1260)p)}{d{\rm cos}\theta}  =
\frac{m^2_p}{8\pi W^2}\frac{|\vec{p_3}^{\rm c.m.}|}{|\vec{p_1}^{\rm
c.m.}|}  \notag \\
&  \times \bigg(\frac{1}{2}
\overline{\sum_{s_2}}\sum_{s_3,s_4}|{\cal M}(\pi^- p \to
a_1^-(1260)p)|^2\bigg) , \label{tcs2}
\end{align}
where $W$ is the invariant mass of the $\pi^- p $ system, whereas,
$\theta$ denotes the scattering angle of the outgoing $a^-_1(1260)$
resonance relative to $\pi^-$ beam direction in the $\rm c.m.$
frame. In the above equation, $\vec{p_1}^{\rm c.m.}$ and
$\vec{p_3}^{\rm c.m.}$ are the 3-momenta of the initial $\pi^-$
meson and the final $a_1(1260)$ mesons,
\begin{eqnarray}
|\vec{p_1}^{\rm c.m.}| &=& \frac{\lambda^{1/2}(W^2, m^2_{\pi^-},
m^2_p)}{2W}, \\
|\vec{p_3}^{\rm c.m.}| &=& \frac{\lambda^{1/2}(W^2, m^2_{a_1},
m^2_p)}{2W},
\end{eqnarray}
where $\lambda(x,y,z)$ is the K\"ahlen or triangle function. We take
$m_p = 938.27$ MeV and $m_{\pi^-} = 139.57$ MeV in this calculation.

In the effective Lagrangian approach, the sum over polarizations and
the Dirac spinors can be easily done thanks to
\begin{eqnarray}
\sum_{s_3} \varepsilon^{\mu}(p_3,s_3) \varepsilon^{\nu
*}(p_3,s_3) &=& -g^{\mu \nu} + \frac{p^{\mu}_3
p^{\nu}_3}{m^2_{a_1}}, \\
\sum_{s_4} \bar{u}(p_4,s_4) u(p_4,s_4) &=& \frac{\Slash p_4 + m_p}{2
m_p}.
\end{eqnarray}

With the formalism and ingredients given above, the calculations of
the differential and total cross sections for $\pi^- p \to \rho^0
\pi^- p$ are straightforward:
\begin{align}
& d\sigma(\pi^- p \to \rho^0 \pi^- p) = \frac{m^2_p}{256 \pi^5
\sqrt{s}} \frac{|\vec p_5^*||\vec p_3|}{\sqrt{(p_1\cdot
p_2)^2-m_1^2m_2^2}} \notag \\
& \times \bigg(\frac{1}{2}\overline{\sum_{s_2}}\sum_{s_3,s_5}|{\cal
M}(\pi^- p \to \rho^0 \pi^- p)|^2\bigg) dM_{\rho\pi} d\Omega^*_5
d\Omega_3, \label{tcs3}
\end{align}
where $\vec p_5^*$ and $\Omega^*_5$ are the three-momentum and solid
angle of the outing $\rho^0$ in the center-of-mass (c.m.) frame of
the final $\pi^- \rho^0$ system, while $\vec p_3$ and $\Omega_3$ are
the three-momentum and solid angle of the final proton in the c.m.
frame of the initial $\pi^- p$ system. In the above equation
$M_{\rho\pi}$ is the invariant mass of the final $\pi^- \rho^0$
two-body system, and $s = (p_1 + p_2)^2$ is the invariant mass
square of the $\pi^- p$ system.

\section{NUMERICAL RESULTS AND DISCUSSION} \label{sec:results}

With the formalism and ingredients given above, the total cross
section versus the beam momentum (${\rm p}_{\rm lab}$~\footnote{The
relation between $W$ (or $s$ for the case of $\pi^- p \to \pi^-
\rho^0 p$ reaction) and ${\rm p}_{\rm lab}$ is: $s = W^2 =
m^2_{\pi^-} + m^2_p + 2m_p \sqrt{m^2_{\pi^-} + {\rm p}_{\rm
lab}^2}$.}) of the $\pi^-$ meson for the $\pi^- p \to a^-_1(1260) p$
reaction is evaluated. The numerical results are shown in
Fig.~\ref{tcs2} for beam energies ${\rm p}_{\rm lab}$ from just
above the production threshold 2.0 to 5.0 GeV together with the
experimental data~\cite{Chung:1968zz,Wohlmut:1970gp} for comparison.
In Fig.~\ref{tcs2}, the dashed, solid, and dotted curves represent
the theoretical results obtained with $\Lambda_{\rho} = 1.4$, 1.5,
and 1.6 GeV, respectively. One can see that the experimental data
can be reproduced with a reasonable value of the cutoff parameter
$\Lambda_\rho = 1.5 \pm 0.1$ GeV. The experimental data from
Ref.~\cite{Chung:1968zz} were measured at ${\rm p}_{\rm lab} = $ 3.2
and 4.2 GeV which can be well reproduced with $\Lambda_{\rho} = $
1.5 GeV. However the experimental data from
Ref.~\cite{Wohlmut:1970gp} at ${\rm p}_{\rm lab} = $ 3.89 GeV is a
few hundred $\mu$b larger than the expected value. More experimental
measurements are needed to complement the limited data in
Refs.~\cite{Chung:1968zz,Wohlmut:1970gp}, and give valuable
information about the mechanism of this reaction.

\begin{figure}[htbp]
\centering
\includegraphics[scale=0.33]{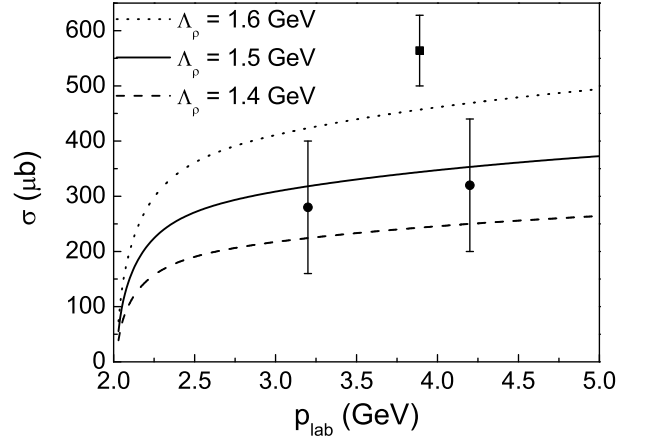}
\caption{Total cross section of $\pi^- p \to a_1^-(1260)p$ reaction
versus the incoming $\pi^-$ beam momentum in the laboratory frame.
The circle data points represent the experimental data from
Ref.~\cite{Chung:1968zz}, while the square point represents the
experimental data from Ref.~\cite{Wohlmut:1970gp}.} \label{tcs2}
\end{figure}

Based on the results of the process of $\pi^- p \to a_1^-(1260)p$,
we investigate the reaction of $\pi^- p \to a_1^-(1260) p \to \rho^0
\pi^- p$. The theoretical calculations of the total cross sections
of this reaction are shown in Fig.~\ref{tcs3}, where we take
$\Lambda_{a_1} = \Lambda_{\rho} = 1.5$ GeV for simplicity. It is
worth to mention that the numerical results are not sensitive to the
value of $\Lambda_{a_1}$. In Fig.~\ref{tcs3}, the dashed, solid, and
dotted curves are obtained with $\Gamma_{a_1} = 250$, 425, and 600
MeV, respectively.

\begin{figure}[htbp]
\centering {\includegraphics[scale=0.33]{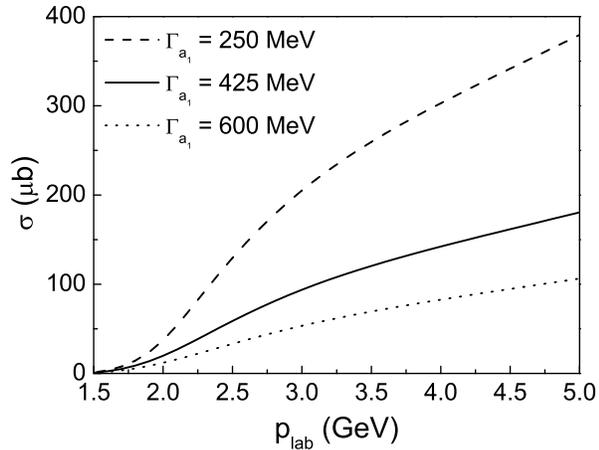}} \caption{Total
cross section of $\pi^- p \to a_1^-(1260) p \to \rho^0 \pi^- p$
reaction versus the $\pi^-$ beam momentum in the laboratory frame.}
\label{tcs3}
\end{figure}

\begin{figure}
\centering {\includegraphics[scale=0.32]{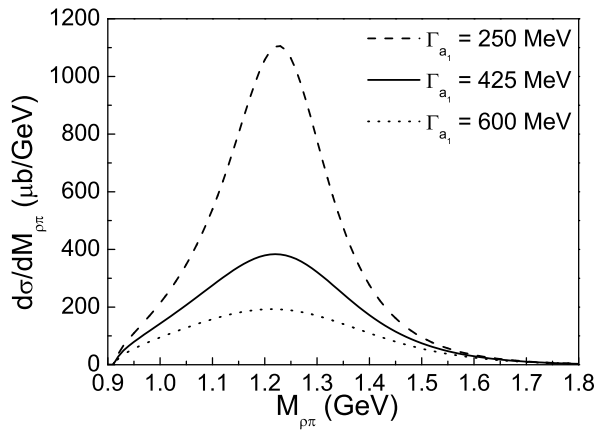}}
\caption{Invariant mass distributions $d\sigma/dM_{\rho\pi}$ of
$\pi^- p \to a_1^-(1260) p \to \rho^0 \pi^- p$ reaction at ${\rm
p}_{\rm lab} = 4$ GeV.} \label{dcs3}
\end{figure}

In addition to the total cross sections of $\pi^- p \to a_1^-(1260)
p \to \rho^0 \pi^- p$ reaction, we calculate also the differential
cross section for this reaction as a function of $M_{\rho \pi}$ at
${\rm p}_{\rm lab} = 4$ GeV. The theoretical results are shown in
Fig.~\ref{dcs3}, where the dashed, solid, and dotted curves are
obtained with $\Gamma_{a_1} = 250$, 425, and 600 MeV, respectively.
The numerical results shown in Figs.~\ref{tcs3} and \ref{dcs3} could
be tested by the future experiments.

\section{SUMMARY}

In this work, we have investigated the $\pi^- p \to a^-_1(1260) p$
and $\pi^- p \to a^-_1(1260) p \to \pi^- \rho^0 p$ reactions near
threshold within an effective Lagrangian approach. The $t$-channel
$\rho^0$ meson exchange process is considered with the assumption
that the $a_1(1260)$ resonance was dynamically generated from the
coupled $K^* \bar K$ and $\rho \pi$ channels, from where we can get
the coupling of $a_1(1260)$ to $\rho \pi$ channel. The total cross
section of $\pi^- p \to a^-_1(1260) p$ is calculated with the
coupling constant of the $a_1(1260)$ to $\rho \pi$ channel obtained
from the chiral unitary theory and a reasonable value of cut off
parameter $\Lambda_\rho$. It is found that the experimental
measurement for the $\pi^- p \to a^-_1(1260) p$ reaction can be
fairly reproduced.

Furthermore, the total and differential cross sections of $\pi^- p
\to a^-_1(1260) p \to \pi^- \rho^0 p$ reaction are also predicted
based on the results of the study of the $\pi^- p \to a^-_1(1260)
p$. Because the width of $a_1(1260)$ resonance has large
uncertainty, we take different values of $\Gamma_{a_1}$ for
comparison. It is expected that our model calculations can be tested
by future experiments.

Finally, we would like to stress that, thanks to the important role
played by the $t$ channel $\rho^0$ exchange in the $\pi^- p \to
a^-_1(1260) p$ reaction, one can reproduce the available
experimental data with a reasonable value of the cut off parameter
in the form factors. The $\pi^- p \to a^-_1(1260) p$ and $\pi^- p
\to a^-_1(1260) p \to \pi^- \rho^0 p$ reactions are important for
the study of the $a_1(1260)$ resonance. More and accurate data for
these reactions will provide valuable information on the reaction
mechanisms and can be used to test our model calculations which
should be tied to the nature of the $a_1(1260)$ state. This work
provides a vision in this direction.

\section*{Acknowledgments}

One of us (C. C.) would like to thank Yin Huang for helpful
discussions. This work is partly supported by the National Natural
Science Foundation of China under Grant Nos. 11475227 and 11475015.
It is also supported by the Youth Innovation Promotion Association
CAS (No. 2016367).

\end{document}